\documentclass[aps,pre,reprint,twocolumn,superscriptaddress,showpacs,floatfix]{revtex4-1}
\usepackage[latin1]{inputenc}
\usepackage{epsfig}
\usepackage{amsfonts}
\usepackage{amssymb}
\usepackage{amsmath}
\usepackage{xcolor}
\usepackage{graphicx}
\usepackage[colorlinks=true,allcolors=blue]{hyperref}
\hypersetup{breaklinks=true}
\usepackage{comment}

\def\ds{{\rm d}s}
\def\da{{\rm d}a}

\begin{document}
\title{Application of Contour Minkowski Tensor and \texorpdfstring{$\mathcal{D}$}{} Statistic to the Planck \texorpdfstring{$E$}{E}-mode data}

\author{Joby P. Kochappan}
\email{jobypk@gmail.com}
\affiliation{School of Physical Sciences, National Institute of Science Education and Research, HBNI, Jatni 752 050, Odisha, India}

\author{Aparajita Sen}
\affiliation{School of Physics, Indian Institute of Science Education and Research Thiruvananthapuram, Thiruvananthapuram 695551, India}
\author{Tuhin Ghosh}
\affiliation{School of Physical Sciences, National Institute of Science Education and Research, HBNI, Jatni 752 050, Odisha, India}
\author{Pravabati Chingangbam}
\affiliation{Indian Institute of Astrophysics, Koramangala 2 Block, Bangalore 560 034, India}
\author{Soumen Basak}
\affiliation{School of Physics, Indian Institute of Science Education and Research Thiruvananthapuram, Thiruvananthapuram 695551, India}

\begin{abstract}
We test the statistical isotropy (SI) of the $E$-mode polarization of the Cosmic Microwave Background (CMB) radiation observed by the Planck satellite using two statistics, namely, the $\alpha$ estimator that is derived from the contour Minkowski Tensor (CMT), and the Directional statistic ($\mathcal{D}$ statistic). The $\alpha$ estimator 
provides information about the alignment of structures and can be used to infer statistical properties such as Gaussianity and SI of random fields. The $\mathcal{D}$ statistic is based on detecting preferred directionality shown by vectors defined by the field. We compute $\alpha$ and $\mathcal{D}$ statistic  for the low resolution component separated SMICA $E$-mode map of CMB polarization, and compare with the values calculated using FFP10 SMICA simulations. We find good agreement between the Planck data and SMICA simulations for both $\alpha$ estimator and $\mathcal{D}$ statistic.
\end{abstract}


\maketitle

\flushbottom
\section{Introduction}
\label{sec:intro}

 The $\Lambda$CDM model has been very successful at explaining the cosmological observations to date, and is currently the most widely accepted model of the Universe \cite{bennett2013nine,akrami2018planck}. In the $\Lambda$CDM model, primordial density fluctuations are assumed to be statistically isotropic (SI). As a consequence the Cosmic Microwave Background (CMB) radiation anisotropies are expected to have the same symmetry property.  Our goal is to test the assumption of SI of CMB polarization using the observations made by the Planck satellite \cite{akrami2018planck,akrami2020planck2}. The CMB polarization field can be split into two components, the curl-free component called the $E$-mode, and the divergence-free component called the $B$-mode. For this work, we focus on the SI test of the CMB $E$-mode polarization, following up our recent work  where we had tested the SI of CMB temperature anisotropy maps \cite{joby2019search}.  
 
It is important to test SI using different approaches that are complementary to each other for obtaining robust constraints. In the literature, various methods can be found for testing the SI of CMB data. Hajian et al.\cite{hajian2004statistical,basak2006statistical,ghosh2007unveiling} formulated the BiPolar Spherical Harmonics (BiPoSH) technique to test the SI of CMB maps. They applied the BiPoSH technique to the WMAP 3-year data, and found no significant violation of SI in the temperature maps, but $\simeq 2\sigma$ deviation from SI \cite{SOURADEEP2006889} in the $E$-mode map. Another technique based in harmonic space is the power tensor method \cite{rath2015testing}, where the eigenvalues of the power tensor constructed from the coefficients of the spherical harmonics contain information about the SI of the field. Using this method, the authors do not find any significant deviations from SI in the temperature maps of the WMAP 3-year data and Planck 2013 data. The authors also applied the power tensor technique to the Planck 2015 polarization data, and found $> 2\sigma$ violation of SI in the multipole range $\ell=40-150$ \cite{rath2018}. More recently, the method of multipole vectors was applied to the Planck 2015 and Planck 2018 temperature data, and the data was found to be in good agreement with the assumption of SI \cite{oliveira2020cmb}. Eriksen et al. \cite{eriksen2007hemispherical} performed a Bayesian analysis on WMAP 3-year data to find $ > 2\sigma$ evidence for the presence of a hemispherical power asymmetry in the temperature maps. An alignment of mild significance was found between the directions of the Planck 2018 CMB temperature and $E$-mode dipolar modulation \cite{akrami2020planck3}, which is again a large scale effect. Considering all these results, further investigation of the SI of the CMB fields at large angular scales is well motivated, and so in this work, we use low resolution Planck 2018 maps.

 Minkowski Tensors (MTs) \cite{mcmullen1997isometry,alesker1999description,hug2008space,schroder2013minkowski,chingangbam2017tensor} carry information on the shapes of the structures. In the context of the CMB, the word structure here refers to hotspots and cold spots defined by iso-field contours. The $\alpha$ estimator, which is derived from one of the MTs, the contour MT (CMT) has been used to test the SI of random fields ~\cite{joby2019search,chingangbam2017tensor,ganesan2017tensor, appleby2018minkowski,appleby2019ensemble}. The CMT is the tensor counterpart of the scalar Minkowski Functional (MF) contour length. The trace of the CMT gives the contour length of the structures. $\alpha$ is sensitive to the kind of SI violation that affects the shapes and relative alignment of the hot spots and cold spots in the field. 
 This method was first applied in~\cite{ganesan2017tensor} to the Planck 2015 data release maps projected stereographically onto a plane. The authors found no significant violation of SI in the CMB temperature field, but obtained $\simeq$ $4\sigma$ deviation for CMB $E$-mode field. Stereographic projection can introduce numerical error in the alignment of structures. A new method for the estimation of the CMT that eliminates projection errors by using field derivatives directly on the sphere, was developed in \cite{appleby2018minkowski}, and was used in \cite{joby2019search} to test SI in the temperature data of the Planck 2018 data release. The polarization part of the  Planck 2018 data has been significantly improved in terms of removal of systematics, foreground modelling and instrumental noise reduction as compared to the Planck 2015 data release \cite{akrami2018planck,akrami2020planck1,aghanim2018planck}. This paper addresses the test of SI of the Planck 2018 polarization data, 
 in particular $E$-mode, using the method of calculation on the sphere.  

 To complement our results from CMTs we use a second test for SI, the Directionality test or the $\mathcal{D}$ statistic \cite{bunn2000preferred}. It is a statistical test which has been devised to measure any preferred directionality over the sky. The test follows a simple formalism, making it numerically inexpensive. It was first applied in  ~\cite{bunn2000preferred} to the COBE-DMR temperature data. Subsequently, the test has also been applied to WMAP and Planck 2018 polarization data in particular to the polarization angle maps in ~\cite{hanson2007directionality} and ~\cite{ghrear2019testing}.

 This article is organized as follows. All the data pre-processing steps that we follow before applying the SI tests have been described in section ~\ref{sec:data}. Section ~\ref{sec:methods} presents the definition of the CMT and $\alpha$, and $\mathcal{D}$ statistic methods.  It also carries a description of how the CMTs can be estimated from pixelated maps of random fields on the sphere. In section ~\ref{sec:noisy}, we discuss how the addition of a SI white noise component to a non-SI (nSI) signal map affects the $\alpha$ and $\mathcal{D}$ values of the resultant map, and demonstrate the sensitivity of our methods to noisy maps. Section \ref{sec:results} presents the main results of this paper, comparisons of the $\alpha$ and $\mathcal{D}$ statistic  estimated from the Planck Spectral Matching Independent Component Analysis (SMICA) $E$-mode map, with those from the FFP10 SMICA simulations. Finally, in section ~\ref{sec:conc}, we draw conclusions based on our results. 


\section{Data}
\label{sec:data} 

In this section, we describe the set of observed data and simulations that we have used for our analysis. We also define the notations used to denote the various sets of maps that we work with.

\subsection{Planck data and mask}
\label{sec:data_mask}
We use the publicly available Planck 2018 data release SMICA $Q$, $U$ maps from the Planck Legacy Archive (PLA). These maps are provided at a beam resolution of $5'$ full-width at half maximum (FWHM) and projected on HEALPix 
pixel resolution of $N_{\rm side}=2048$ \cite{akrami2020planck2}. These maps combine multi-frequency sky observations by the Low Frequency Instrument (LFI, $30-70$\,GHz) and the High Frequency Instrument (HFI, $100-857$\,GHz). The HFI has an angular resolution of $\approx 5'-10'$ FWHM, while the LFI has a resolution of $\approx 13'-33'$ FWHM. As Stokes $Q$ and $U$ maps are spin-2 quantities for which the morphological properties are strongly affected by masking~\cite{chingangbam2017minkowski}, we choose to convert them into $E$ and $B$-mode maps, which are invariant under rotation. We use ``anafast" routine of HEALPix to transform the full sky Stokes $Q$ and $U$ maps into spherical harmonic coefficients $E_{\ell m}$s and $B_{\ell m}$s. We first deconvolve the beam response of the original SMICA polarization maps from the $E_{\ell m}$s and then convolve the $E_{\ell m}$s with a Gaussian beam of $1^{\circ}$ FWHM in harmonic space. Then we use ``synfast" routine of HEALPix to transform back the $E_{\ell m}$s into the $E$ mode map at $N_{\rm side}=128$. The smoothing process is required for reducing the level of noise in the data, since the reconstructed SMICA $E$-mode map is noisy at the original $5'$ resolution (see figure 16 of \cite{akrami2020planck2}). The pixel resolution of the smoothed $E$-mode map is chosen in such a way that the effective beam falls over three pixels. 
We will refer to the smoothed SMICA $E$-mode map as the data $E$ map.

\begin{figure}
\centering
\includegraphics[width=1.0\linewidth]{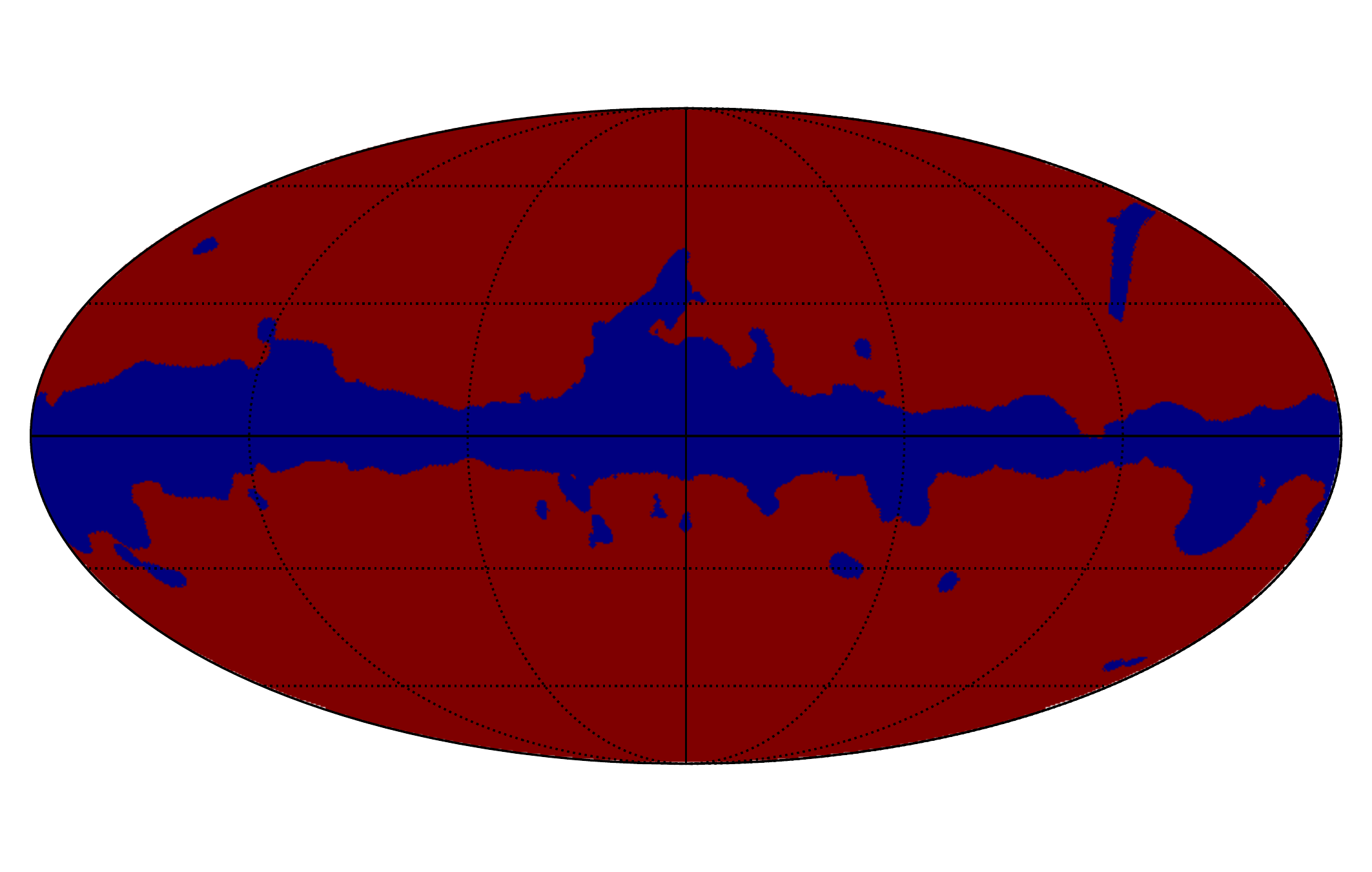}
\caption{The binary mask used in our analysis at $N_{\rm side}=128$ has an effective sky fraction of 78\%.}
\label{fig:mask}
\end{figure}

For the analysis of Planck polarization data, the recommended mask is the common polarization mask, which is available at a resolution of $N_{\rm side}=2048$ and has an effective sky fraction of $78\%$ \cite{akrami2020planck2}. We downgrade the Galactic component of the polarization mask to $N_{\rm side}=128$ and we refer to it as the P78 mask. The P78 mask has a sky fraction of 78\%. The binary P78 mask is shown in figure ~\ref{fig:mask}.

\subsection{Simulations}
\label{sec:data_sims}

We use the Full Focal Plane (FFP10) SMICA $Q$ and $U$ simulated maps to make a comparison with the results obtained from the  data $E$ map. The SMICA pipeline processed 1000 CMB-only simulations and 300 noise-only simulations separately by applying the same frequency weights as derived from the real Planck data \cite{akrami2020planck2}. The noise simulations include the instrumental noise and residual systematics processed through end-to-end simulations. It is very important to account for the residual systematic effects to test the SI violation of the data $E$ map at large angular scales.  For our purpose, we use the linear combination of the first 300 SMICA CMB-only and noise-only simulations. These simulations have the same beam resolution as the SMICA data $Q$, $U$ maps. We perform the same post-processing of the CMB+noise simulations as we do on the Planck data to produce the smoothed simulated maps at $1^{\circ}$ FWHM beam resolution and $N_{\rm side}=128$. We refer to these as the SMICA simulations. Then we mask all the maps with P78 mask before applying our SI estimators. 

\begin{figure}
\centering
\includegraphics[width=1.0\linewidth]{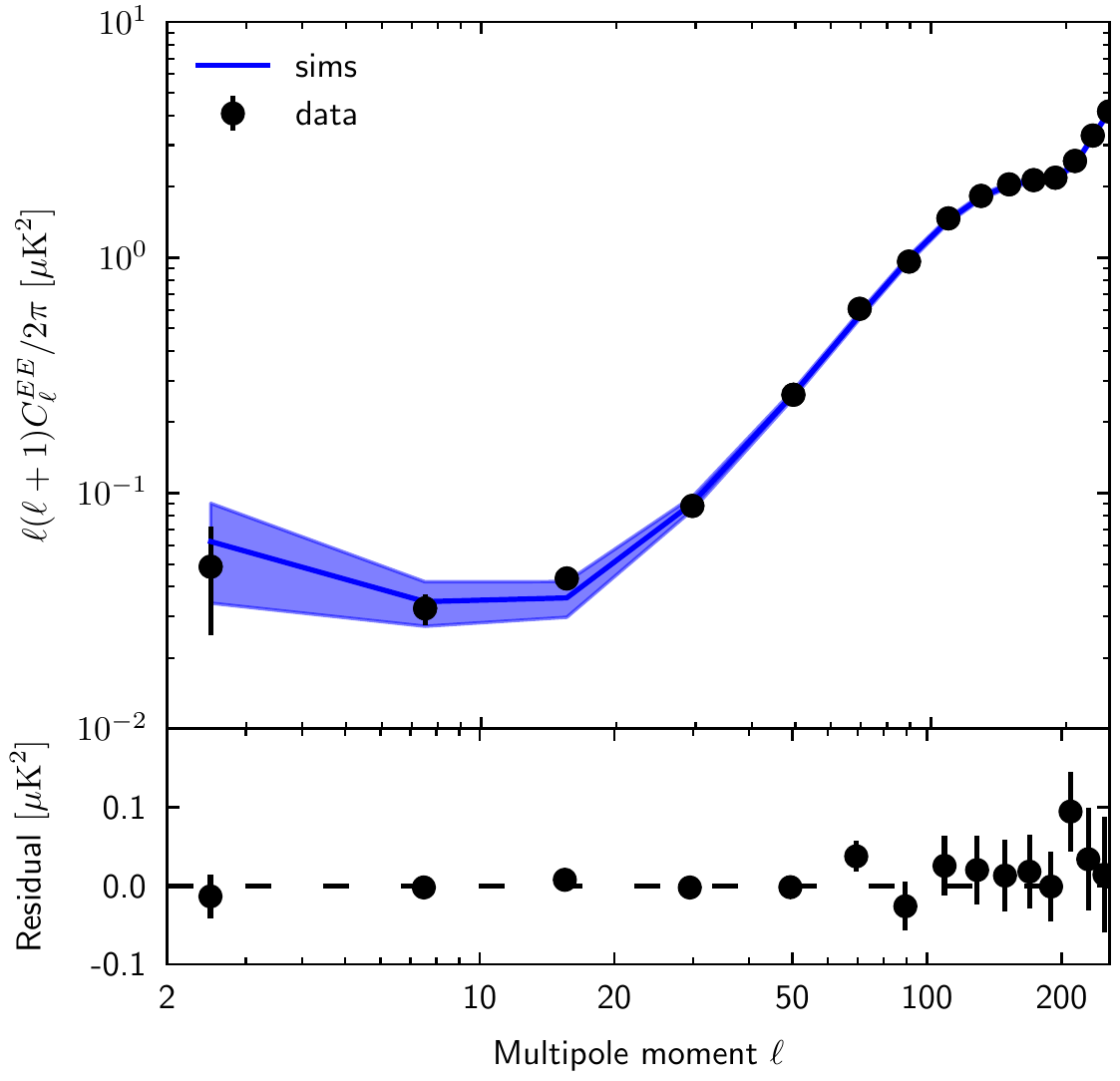}
\caption{{\em Top panel}: Comparison of the full sky $EE$ power spectra estimated from the data $E$ map (black points) and mean power spectrum from 300 SMICA simulations (blue line).  All the power spectra are computed from smoothed downgraded $E$ map ($1^{\circ}$ FWHM beam resolution and $N_{\rm side}=128$) with apodized P78 mask applied. {\em Bottom panel}: The residual $E$-mode power spectrum after subtracting the mean simulated power spectrum, from SMICA simulations from the Planck $EE$ power spectrum along with 1-$\sigma$ error bars.}
\label{fig:power_spectra}
\end{figure}

\subsection{Level of noise in the Planck \texorpdfstring{$E$}{E} map}
\label{sec:noise_level}

In this section, we compute the $EE$ power spectrum of the data and simulations over the masked sky to estimate the level of noise in the data $E$ map. For the power spectrum estimation, we apodized the P78 mask to avoid power leakage due to sharp mask boundaries. The P78 mask is convolved with $5^\circ$ FWHM Gaussian beam to produce the apodized P78 mask. The mask apodization is chosen in such a way that it smoothly goes to zero towards the mask edges. The effective sky fraction after applying apodization becomes $71\%$. We apply the apodized P78 mask to the data $E$ map at $N_{\rm side}$=128 and compute the pseudo $EE$ power spectrum. We then use the Xpol package \cite{xpol} to estimate the full-sky $EE$ power spectrum, corrected for the masking, beam and pixel window effects. The same procedure has been applied to the SMICA simulations to compute the simulated $EE$ power spectrum for a given sky realization. For the simulations, we calculate the mean $EE$ power spectrum by averaging the power spectra obtained from 300 sky realizations. In the top panel of figure ~\ref{fig:power_spectra}, we compare the scaled power spectrum, $\ell(\ell+1)C_{\ell}^{EE}/2\pi$, where $C_{\ell}^{EE}$ is the $EE$ power spectrum, of the  Planck data with  the mean from SMICA simulations. In the bottom panel of figure ~\ref{fig:power_spectra}, we show the residual power spectrum, which represents the difference in the power spectrum of the data from the mean of the SMICA simulations, with 1-$\sigma$ error bars from the simulations. Since maps of the foreground residuals are not available, and the SMICA simulations and noise simulations are consistent with the data at the scales we are probing, we do not consider the effect of foreground residuals in our analysis. We define the signal to noise ratio ($\gamma$) of the data $E$ map as the ratio of the root mean square (rms) of the SMICA CMB-only and noise-only simulations in the pixel space. For the data $E$ map, $\gamma=1.04$.


\section{Quantifying the SI of CMB maps}\label{sec:methods}
In this section, we describe the two statistical estimators that we use to test the SI of the data $E$ map, namely, $\alpha$ and $\mathcal{D}$ statistic. We start with the definitions of the two statistics and then discuss how they capture the SI information of a given random field.

\subsection{Contour Minkowski Tensor}
\label{sec:mt_on_sphere}

First, we give a brief overview of the CMT, and define the $\alpha$ estimator, which is our primary tool for testing SI. This technique for testing the SI of a random field defined on the surface of a sphere was developed in \cite{chingangbam2017tensor}. 
Let $C$ be a closed curve on the unit sphere. The CMT associated with this curve, denoted by $\mathcal{W}_1$, is defined as,
\begin{equation}
    \mathcal{W}_1 = \frac{1}{4}\int_C \hat{T} \otimes \hat{T} \, \ds ,
    \label{eqn:cmt}
\end{equation}
where the integral is over $C$, $\hat{T}$ denotes the unit tangent vector to the curve at each point, $\otimes$ denotes the symmetric tensor product of two vectors, and $\ds$ denotes the infinitesimal arc length line element along the curve. The ratio of the two eigenvalues of $\mathcal{W}_1$ represent the isotropy of the curve $C$. If $C$ is isotropic, then the two eigenvalues are equal. 

For a smooth random field, denoted by $u$, excursion sets consist of points on the sphere where the field has values higher than some chosen threshold value ($\nu$). The boundaries of the excursion sets, indexed by $\nu$, form closed curves. For multiple such curves, $\mathcal{W}_1$ can be obtained as the sum over all the curves. To compute $\mathcal{W}_1$ numerically for excursion set boundaries at each $\nu$ on the sphere, we can convert the line integral to an area integral using a suitable Jacobian \cite{schmalzing1998minkowski}, and express it in terms of the field and its first  derivative components  $u_{;\theta}$ and $u_{;\phi}$, where $\theta,\phi$ are spherical coordinates on the sphere, as \cite{chingangbam2017tensor,joby2019search},
\begin{equation}
    \mathcal{W}_1 = \frac{1}{16\pi}\int_{S^2} \da \, \delta \left( u-\nu \right) \, \frac{1}{\left|\nabla u\right|} \, \begin{pmatrix}u_{;\phi}^{2} & -u_{;\theta}u_{;\phi}\\
-u_{;\theta}u_{;\phi} & u_{;\theta}^{2} \end{pmatrix},
    \label{eqn:cmt_area}
\end{equation}
where $S^2$ indicates that the integral is over the unit sphere, $\da$ is the area element on the sphere, $\delta$ is the Dirac delta function, and $\nabla$ denotes covariant derivative on the sphere. Choosing to label the eigenvalues of $\mathcal{W}_1$ as $\Lambda_1$ and $\Lambda_2$ such that $\Lambda_1 \leq \Lambda_2$, the alignment parameter, which we refer to as the $\alpha$ estimator, is defined as the ratio of the two eigenvalues \cite{chingangbam2017tensor},
\begin{equation}
    \alpha = \frac{\Lambda_1}{\Lambda_2} \ . \label{eqn:alpha}
\end{equation} 

By construction, the values of $\alpha$ range between 0 and 1. For all the structures corresponding to a selected threshold value of the field, the locus curve is defined in \cite{chingangbam2017minkowski} using the mean radial distances of the curves from their centroid, when the curves are stacked together. $\alpha$ represents the isotropy of the locus curve formed this way. $\alpha$ gives a measure of the alignment of the structures in the level sets, which reflects the SI of the random field. If the field is SI, then the value of $\alpha$ will be close to unity. The value of $\alpha$ will  shift towards zero for non-SI (nSI) field. As we work with the field rescaled by its standard deviation, $\nu$ represents the rescaled threshold. The calculation of $\alpha$ using equation ~\ref{eqn:alpha} is quite accurate and the numerical error arising from the discrete $\nu$ binning approximation of the $\delta$ function is small (see the discussion of eqn.~2.9 in ~\cite{goyal2020morphology}).

\begin{figure}[h!]
    \centering
    \includegraphics[width=1.0\linewidth]{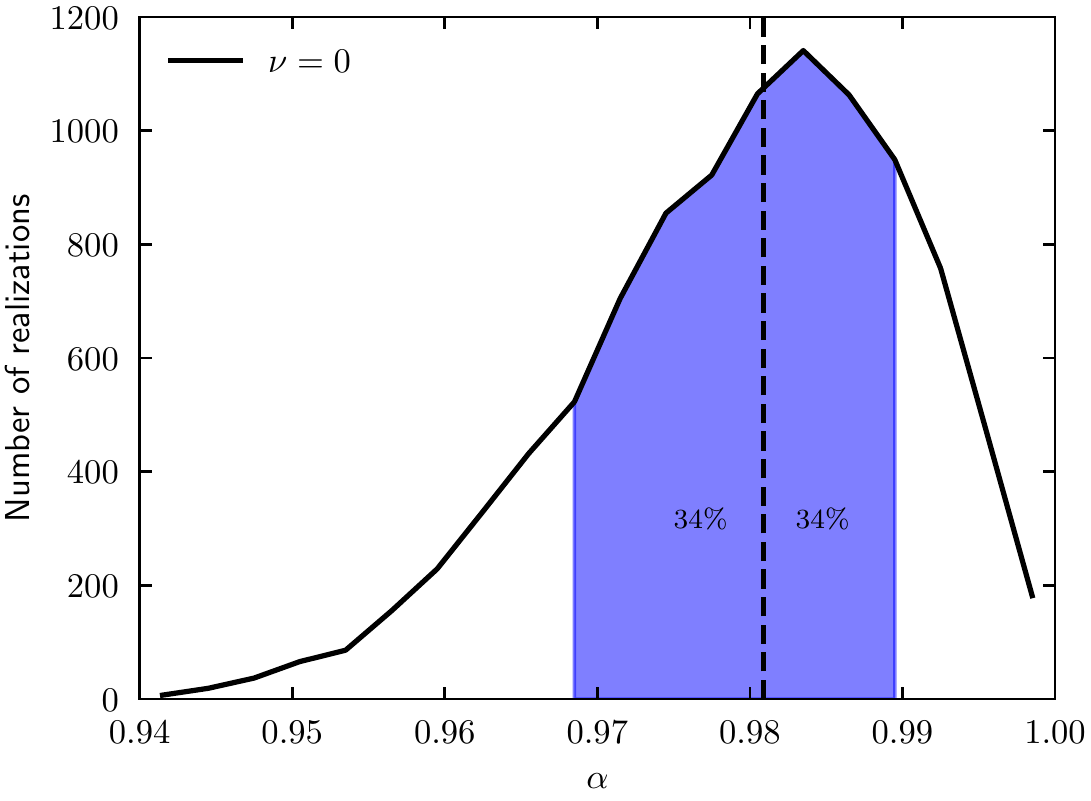}\\
    \includegraphics[width=1.0\linewidth]{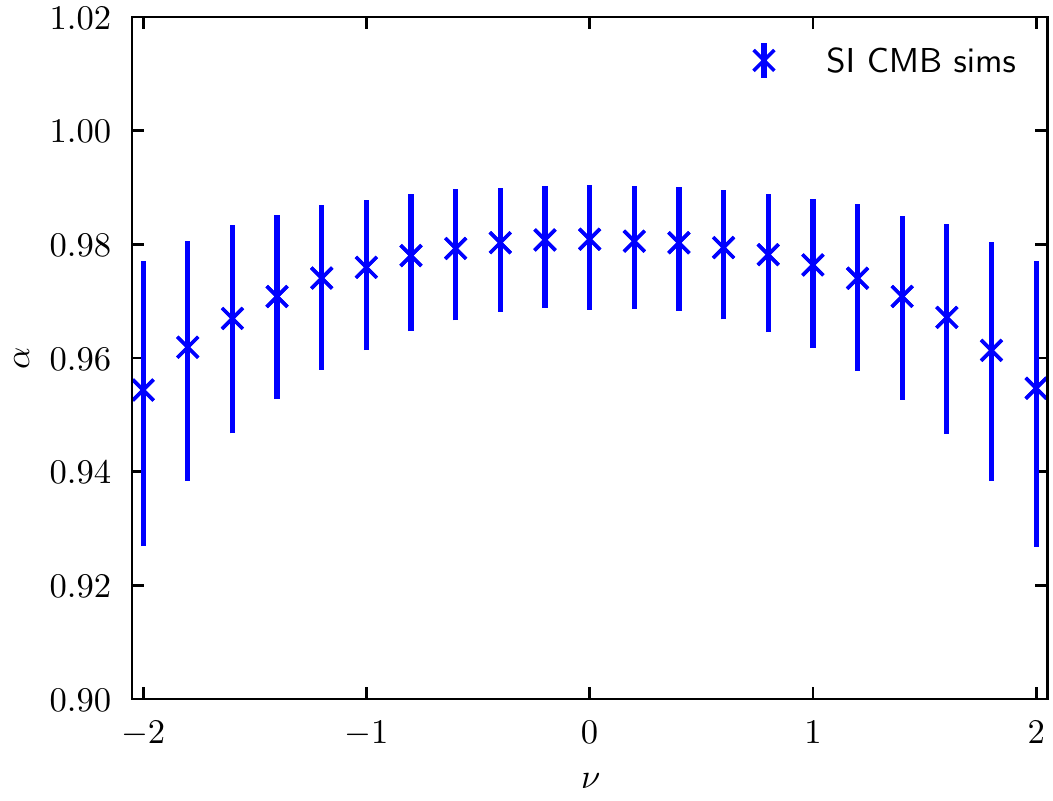}
    \caption{{\it Top panel}: A histogram of the $\alpha$ values obtained from 10000 SI CMB simulations for threshold value $\nu$=0. {\it Bottom panel}: Median of $\alpha$ values and 68\% limits, obtained from 10000 SI simulations with the Planck best-fit $\Lambda CDM$ power spectrum (blue points).}
    \label{fig:a_dist}
\end{figure}

We demonstrate $\alpha$ values estimated from SI maps in figure \ref{fig:a_dist}. The top panel shows the distribution of $\alpha$ from a set of 10000 SI CMB $E$-mode simulations ($N_{\rm side}=128$ and beam resolution=$1^{\circ}$) drawn from the Planck best-fit $\Lambda CDM$ model \cite{cosmo_params} for the threshold value $\nu=0$. The probability distribution function of $\alpha$ follows the Beta distribution. There is a mild dependency of $\alpha$ on the input power spectrum of the SI signal ~\cite{goyal2020morphology}. The bottom panel of figure \ref{fig:a_dist} shows the median $\alpha$ and error bars representing the 68\% limits, calculated from 10000 SI CMB $E$-mode simulations over the threshold range between $-2$ and 2. The error bars are asymmetric and to estimate them, we first sort the values of $\alpha$ at each threshold in increasing order. Next, we multiply the total number of simulations by 0.34 (34\%) and get the index of the simulations, above and below the median, representing the upper and lower limits, respectively. This is the method we follow to estimate the error bars on $\alpha$ throughout this paper.

\subsection{Directionality Test}
\label{sec:D-Stat Theory}
In this section we will briefly describe the basic concepts behind the preferred directionality test ~\cite{bunn2000preferred, hanson2007directionality, ghrear2019testing}. 
It is a statistical test defined in the pixel domain to detect any preferred directionality present in the observed CMB signal. To achieve this goal, a vector at each point in the sky is defined to capture the directional properties of the field. For a scalar field $u$ the vector can simply be the gradient of $u$. The alignment of these gradients towards any particular direction is quantified by its projection along that direction of the sky. Mathematically, this can be implemented by defining a function $f$ for each direction $\hat{\boldsymbol{n}}$ in the sky,
\begin{equation}
    f(\hat{\boldsymbol{n}})=\sum_{p=1}^{N_{\rm pix}}w_{p}\left(\hat{\boldsymbol{n}}.\vec{\nabla}u_p\right)^{2},
    \label{eq: f_n_cap}
\end{equation}
where, $p$ stands for the pixel index of the map and $N_{\rm pix}$ is the total number of pixels on the map. $w_p$ are the weights for each pixel on the map and are used to cancel out any false directional signal due to masking \cite{bunn2000preferred}. Any anomalously large or small value of $f(\hat{\boldsymbol{n}})$ will indicate the presence of a directionality towards $\boldsymbol{\hat{n}}$ or $-\boldsymbol{\hat{n}}$ and if there is no preferred directionality present in the sky, all the values of $f(\hat{\boldsymbol{n}})$ should be close to each other. This information is captured by the $\mathcal{D}$ statistic which is defined as,
\begin{equation}
    \mathcal{D}=\frac{{\rm max} [f(\hat{\boldsymbol{n}})]}{{\rm min} [f(\hat{\boldsymbol{n}})]}.
    \label{eq:D-ratio}   
\end{equation}
$\mathcal{D}$ takes real positive values greater than or equal to unity. For a SI map the values of $f(\hat{\boldsymbol{n}})$ for different directions are expected to be close to each other as no particular direction is favoured and hence  $\mathcal{D}$ will be close to unity. So, any nSI feature present in the map will manifest itself as deviations of $\mathcal{D}$ from unity. We compare the $\mathcal{D}$ value of the data with the corresponding values obtained from SI simulations to detect any violation of SI.

\section{Sensitivity of \texorpdfstring{$\alpha$}{} estimator and \texorpdfstring{$\mathcal{D}$}{} statistic}
\label{sec:noisy}

The $\alpha$ estimator and $\mathcal{D}$ statistic are designed for testing the SI of signal dominated maps. The SI property of noisy maps can be significantly affected by the properties of the noise. For example, suppose the CMB $E$ mode signal is nSI, but a dominant SI noise could make the resultant map SI, and vice versa. Since the data $E$ map is known to be noisy at the scales we are interested in, we first check the sensitivity of $\alpha$ and $\mathcal{D}$ to pick up the nSI property of the signal in presence of SI noise. For this purpose, we make a toy model of nSI map using the foreground model $E$ mode map at 353 GHz. In the 353 GHz band, thermal dust is the dominating component and so we use the thermal dust map for our toy model. We start with the FFP10 thermal dust template Stokes $Q$ and $U$ maps at 353 GHz, which are publicly available on PLA. We first convert the full sky $Q$ and $U$ maps to spherical harmonic coefficients $E_{\ell m}$ and $B_{\ell m}$. We then deconvolve the input map pixel window and beam functions from the $E_{\ell m}$s and apply a bandpass filter $f_\ell$ given by,
\begin{equation}
  \centering
   f_{\ell} = \left\{ \begin{tabular}{cl}
        0 & \text{if }  $\ell$ $<$ 35\\
         & \\
        $\cos^2{\left(\frac{\pi}{2} \cdot \frac{\left(40 - \ell\right)}{5} \right)}$ & \text{if } 35 $\leq$ $\ell$ $\leq$ 40, \\
         & \\
        1 & \text{if } $\ell$ $>$ 40.
   \end{tabular} \right.
\end{equation}
Next, we use the HEALPix routine ``smoothing'' to apply a Gaussian smoothing of $1^{\circ}$ FWHM in harmonic space, and reproject the $E$ mode map at $N_{\rm side}$=128. To this dust $E$ map ($d$), we add varying levels of SI white noise ($n$), to produce the noisy nSI maps. We choose to work with $1^{\circ}$ FWHM Gaussian smoothed nSI maps as they mimic the smoothing present in the data $E$ map.

\begin{figure}
    \centering
    \includegraphics[width=1.0\linewidth]{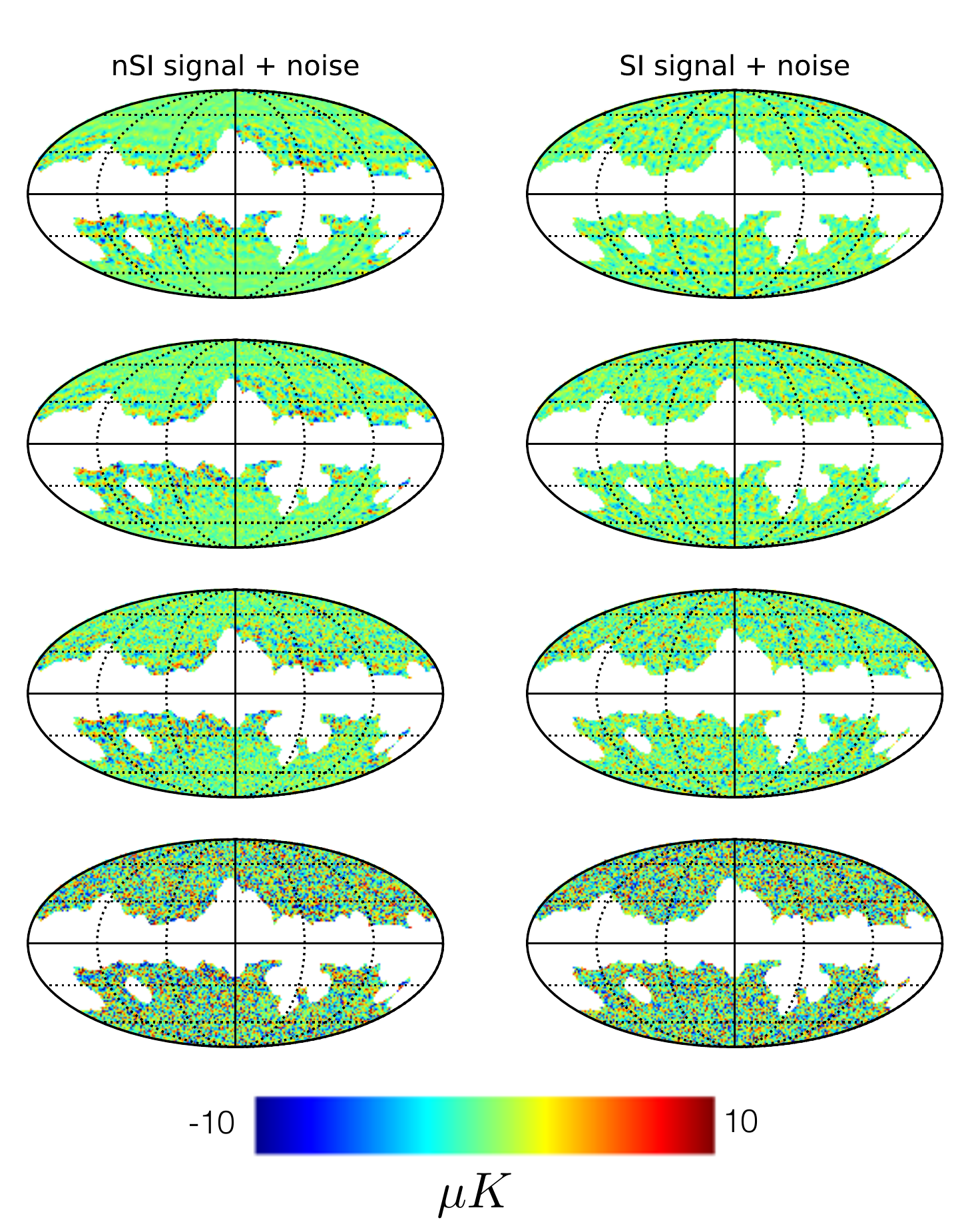}
    
    \caption{Visualization of the noisy dust maps (\textit{left panels}) and corresponding SI simulations (\textit{right panels}). From top to bottom, the panels display the maps with only dust, and those with added noise having $\gamma$ ratios of 2, 1 and 0.5, respectively.}
    \label{fig:maps_fg+wnoise}
\end{figure}

\begin{figure}
    \centering
    {\includegraphics[width=1.0\linewidth]{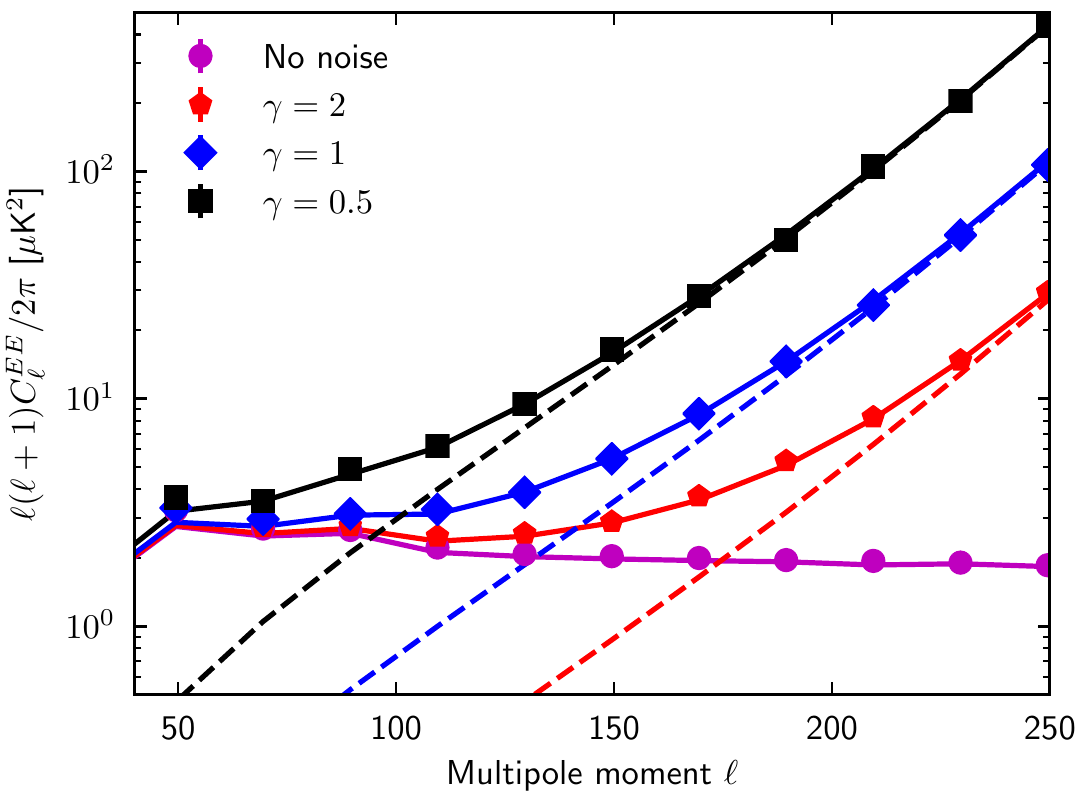}}
    \caption{Beam deconvolved $EE$ power spectra for the noisy dust maps (\textit{points}), mean of 1000 SI simulations (\textit{solid lines}) and added SI white noise (\textit{dashed lines}). The maps having only the dust component and those with added noise having $\gamma$ ratios of 2, 1 and 0.5, are shown in \textit{purple}, \textit{red}, \textit{blue} and \textit{black} colours, respectively.}
    \label{fig:power_fg+wnoise}
\end{figure}

The rms amplitude of the dust map is dominated by regions close to the Galactic plane having high standard deviation values. For the calculations in this section, we use a mask which excludes the regions having standard deviation (computed over lower $N_{\rm side}=16$) higher than a selected threshold value of $6\, \mu$K. We will refer to this mask as the dust mask, which has a sky  fraction  of  $60\%$.
We add different levels of noise to the input nSI map by varying the ratio of rms amplitude of the $d$ map and $n$ map in the pixel space defined as,
\begin{equation}
    \gamma = \frac{\sigma_{d} (\rm nSI)}{\sigma_{n} (\rm SI)} \ .
\end{equation} 
For our analysis, we choose nSI maps having $\gamma$ values, $\gamma$= 2, 1 and 0.5, in addition to the case where no noise is added. Figure \ref{fig:maps_fg+wnoise}  displays the noisy nSI maps alongside an SI realization with the same underlying $EE$ power spectrum, for the dust-only case, and with the three other selected $\gamma$ values. We note that the addition of noise leads to an increased number of small scale grainy structures, while leaving the large scale patterns intact. We then mask these maps with $5^{\circ}$ FWHM Gaussian apodized dust mask, and use the Xpol package \cite{xpol} to estimate the full-sky $EE$ power spectra. With these power spectra as the inputs in HEALPix, we generate SI simulations for each of the selected levels of $\gamma$. Figure ~\ref{fig:power_fg+wnoise} displays the power spectra of the noisy nSI dust maps (points), mean from 1000 corresponding SI simulations (solid lines), and added noise (dashed lines) for the three different $\gamma$  ratios. The power spectra extracted from the SI maps, and from noisy nSI dust maps, overlap. As expected, we can see that with increasing levels of added noise, the power spectra become dominated by noise at small scales.

 In the following subsections we compare the results for the $\alpha$ estimator and $\mathcal{D}$ statistic using the noisy nSI maps with the corresponding SI simulations, which have the same $EE$ power spectrum. In each case, the noisy nSI maps, and the SI simulations, are masked with the dust mask before applying the $\alpha$ estimator and $\mathcal{D}$ statistic techniques. The goal of this exercise is to demonstrate the sensitivity of the $\alpha$ estimator and the $\mathcal{D}$ statistic.

\subsection{\texorpdfstring{$\alpha$}{} estimator}
\label{sec:noisy_alpha}
\begin{figure}
    \centering
    \includegraphics[width=1.0\linewidth]{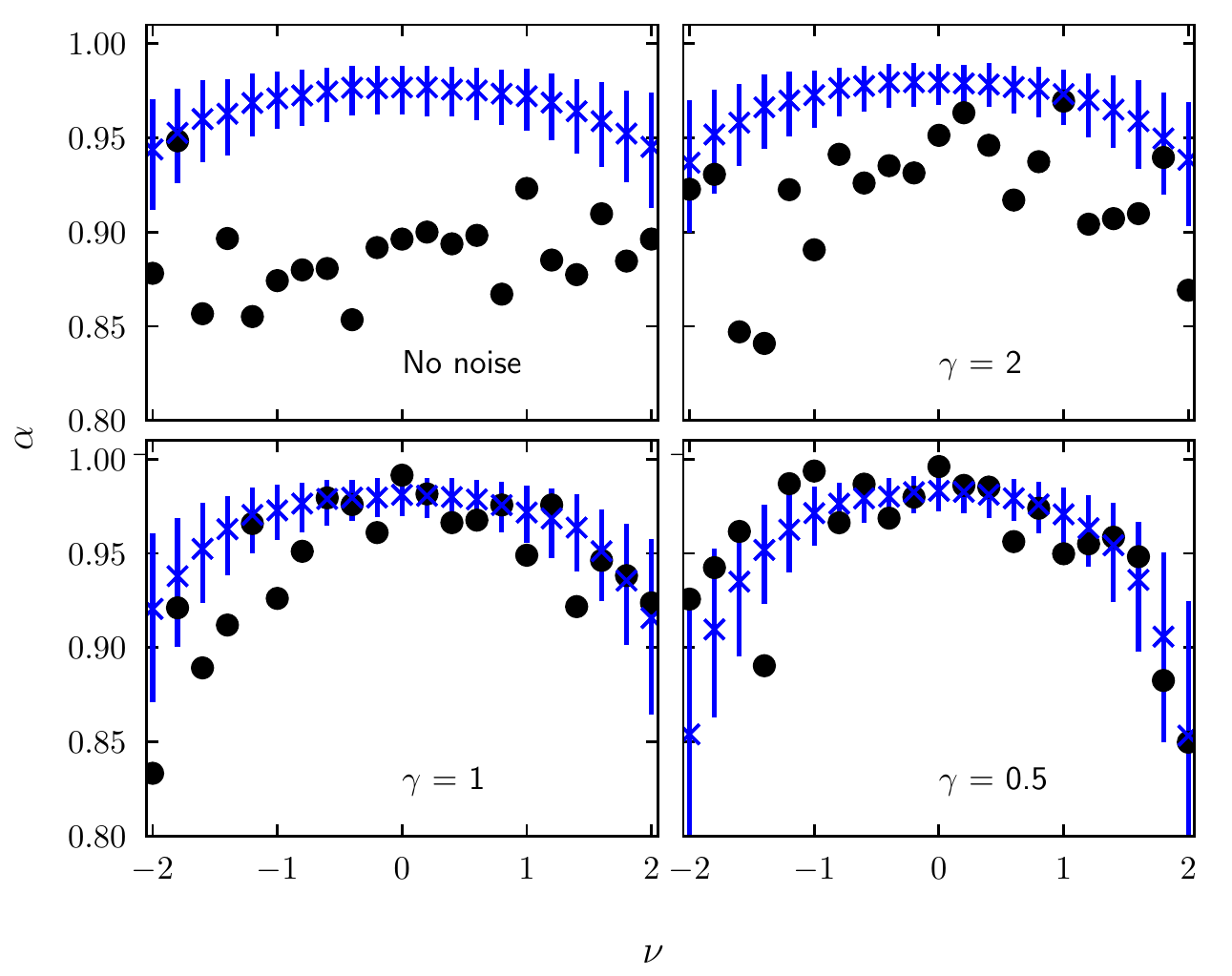}
    \caption{Comparison of $\alpha$ from noisy dust maps (black points),  median $\alpha$ and error bars representing 68\% limits (blue crosses) from 1000 SI simulations having the same $EE$ power spectra, for the no noise map (\textit{top left}), and added noise with $\gamma$ ratios of 2 (\textit{top right}), 1 (\textit{bottom left}) and 0.5 (\textit{bottom right}) panels respectively.}
    \label{fig:alpha_fg+wnoise}
\end{figure}

Figure \ref{fig:alpha_fg+wnoise} presents the results for the $\alpha$ estimator applied on the noisy nSI dust maps with three different $\gamma$ ratios of  2, 1 and 0.5 along with the no noise case. In each panel, the black points represent $\alpha$ values for the noisy nSI dust map, while the blue crosses with error bars represent the median and 68\% limits from 1000 SI simulations generated from the same underlying $EE$ power spectrum as the noisy nSI dust map computed over the apodized dust mask. The addition of noise increases the $\alpha$ values of the resultant map and makes it effectively SI, even though the original signal of our interest is nSI. The value of $\alpha$ at any threshold depends on the number of structures present in the field at that threshold. Due to this, and combined with the fact that we have a single realization of the observed CMB sky, the values of $\alpha$ for individual thresholds fluctuate about their expectation values. Thus, combining the information in $\alpha$ values over a threshold range rather than looking at individual thresholds provides a much better statistic for inferring the property of SI. The spread of $\alpha$ about the corresponding median value is relatively low (high statistical significance) for thresholds close to 0, and it increases as we move away from 0 on either side. This is because at large $\left|\nu\right|$, most of the structures are isolated local maxima or minima, and get easily washed out due to downgrading \cite{joby2019search}. For our analysis, we choose the threshold range $\nu = -2$ to $+2$. We compute the correlation between different threshold values from $-2$ to $+2$ using SI simulations and SMICA simulations, and find that the $\alpha$ values are uncorrelated between two nearby threshold values in both cases. 

To quantify the consistency between the data and the simulations, we use the $p$-value statistic, which requires no assumptions regarding the shape of the probability density distribution of $\alpha$. For $\alpha$, the $p$-value at each threshold is defined as the probability of obtaining $\alpha$ values lower than the data at that threshold, based on the simulations. Here, the simulations refer to the 1000 SI simulations with the same $EE$ power spectra as the noisy dust maps. We use the ``combine\_pvalues'' routine from the scipy package \cite{scipy}, with the Fisher method \cite{fisher1932}, to obtain a single combined $p$-value for the noisy nSI dust map, in the threshold range $-2$ to $+2$. The Fisher method, constructs a test statistic using the logarithms of each of the $p$-values. When the individual $p$-values are independent of each other, the test statistic has a $\chi^2$ distribution and a single $p$-value for the distribution can be estimated. In our case, the nearby thresholds are not correlated and hence the Fisher method is suitable for providing a combined estimate of the $p$-value. For no noise case and $\gamma$=2, the noisy dust maps strongly disagree with the corresponding SI simulations. For $\gamma$=1, the combined $p$-value for the noisy dust map based on the SI simulations is 0.0002, which means that the corresponding noisy dust maps are not consistent with SI. Although the black points are consistent with the blue error bars for a lot of the thresholds, it is important to note that the black points are consistently lower than the blue crosses for almost all thresholds. This leads to small $p$-values at all the thresholds resulting in a very small combined $p$-value and inconsistency between the noisy dust map and SI simulations, where a simple comparison with the error bars would suggest that the noisy dust map is consistent with the SI simulations. The combined $p$-value captures the information regarding the data points being systematically lower than the simulations, which would otherwise be missed by a simple comparison of error bars. As we increase the level of noise in the map (or decrease the $\gamma$ ratio) to $\gamma$=0.5, the combined $p$-value increases to 0.41, indicating that the noisy nSI dust map becomes consistent with the SI simulations.

\subsection{\texorpdfstring{$\mathcal{D}$}{D} statistic}
\label{ssec:D-Stat Sensitivity}
\begin{figure}
    \centering
    \includegraphics[width=1.0\linewidth]{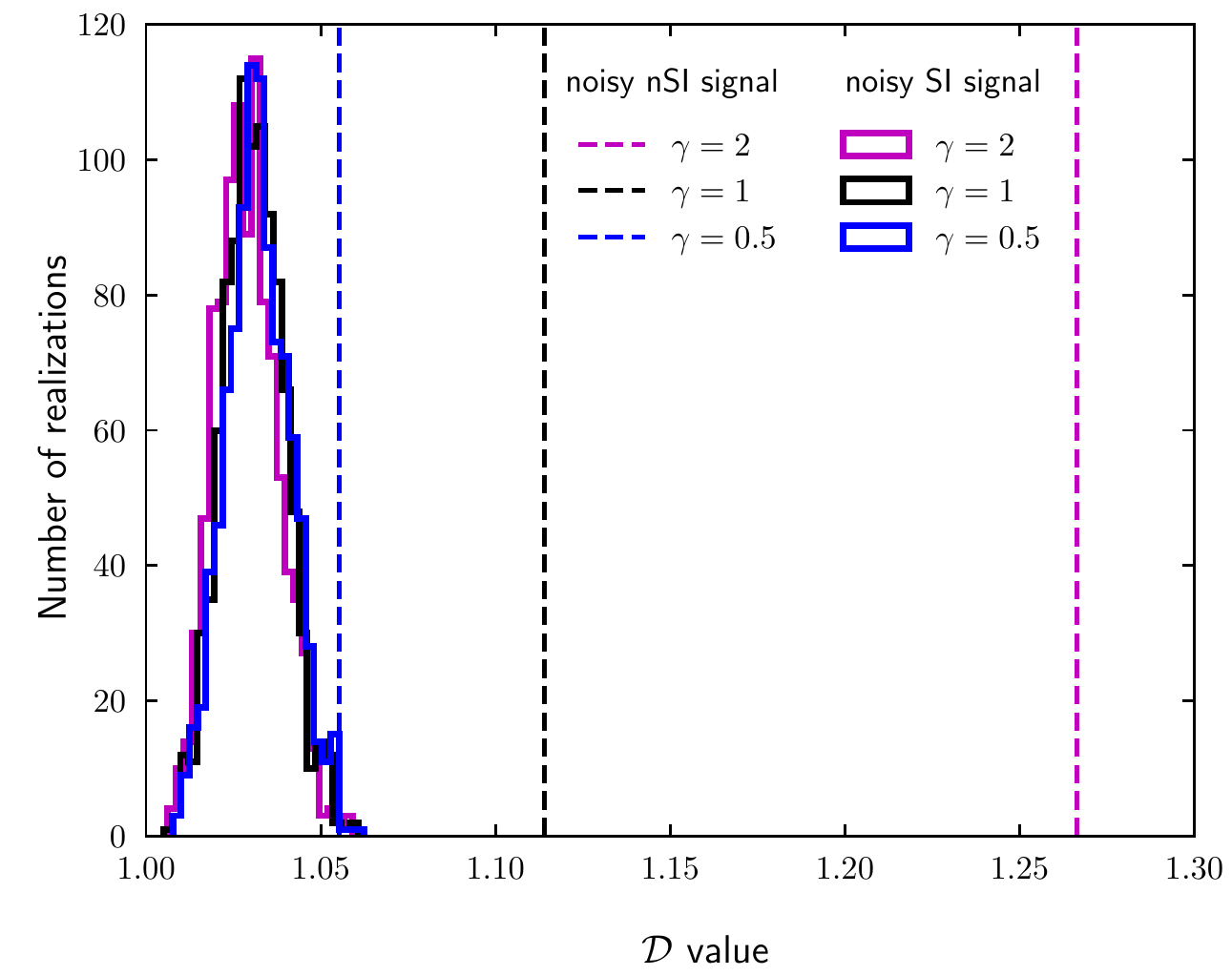}
    \caption{Comparison of $\mathcal D$ from noisy dust $E$ maps and the corresponding 1000 SI simulations for three values of $\gamma$: 2 (magenta), 1 (black) and 0.5 (blue).}
    \label{fig:353fg+wnoise}
\end{figure}

We analyze the sensitivity of $\mathcal{D}$ statistic on the noisy dust nSI $E$ maps by comparing the $\mathcal{D}$ value obtained from noisy dust maps with the corresponding values obtained from 1000 SI simulations having the same $EE$ power spectrum.
For a statistically significant detection of SI violation, we compute the $p$-value from the noisy dust maps. For $\mathcal{D}$ statistic, the $p$-value is defined as the probability of obtaining $\mathcal{D}$ values higher than the data, based on the simulations. In this case, the simulations are the corresponding 1000 SI simulations. Figure \ref{fig:353fg+wnoise} presents the results from $\mathcal{D}$ statistic for $\gamma$= 0.5, 1 and 2. The dust maps with no added noise and those with added noise having $\gamma$=2 and 1, strongly disagree with the SI simulations. At $\gamma$=0.5, the $p$-value is 0.009, and the noisy dust map is statistically consistent with SI, within the 99.7\% limits. The small $p$-value for $\mathcal{D}$ at $\gamma$=0.5 suggests that $\mathcal{D}$ is more sensitive than $\alpha$ at detecting the type of nSI signal present in our toy model map.

From these results, we demonstrate two salient features of $\mathcal{D}$ statistic. First, $\mathcal{D}$ is quite robust in detecting low signal-to-noise nSI signal even in the presence of SI white noise. However, $\mathcal{D}$ statistic sensitivity towards detecting the SI violation fails for very low $\gamma$ value ($\gamma < 0.5$). Secondly, as the level of the SI noise is increased in the original dust nSI map, the noisy maps become consistent with SI. We expect the $\mathcal{D}$ value closer to unity for SI signal and much higher values for nSI signal. For SI simulations, we find that the distribution of $\mathcal{D}$ is almost identical for different $\gamma$ values. However, for the noisy nSI dust maps, the $\mathcal{D}$ value, decreases with the decrease in $\gamma$. In no noise case, the $\mathcal{D}$ value from the nSI dust map is 1.64, making it difficult to plot alongside the $\mathcal{D}$ values from the other maps.

Both our techniques, $\alpha$ estimator and $\mathcal{D}$ statistic are sensitive enough to pick up the nSI of the signal from noisy maps up to $\gamma$=1.
For the data $E$ map, the value of $\gamma$ is marginally greater than 1 over the masked region. Based on the toy model, we expect to detect the SI violation (if any) present in the CMB $E$-mode polarization using $\alpha$ and $\mathcal{D}$.

\section{Results}
\label{sec:results}
We carry out our stated goal of testing SI using data $E$ map in this section. We compare the values of $\alpha$ and $\mathcal{D}$ statistics obtained from the Planck data with the SMICA simulations.

\begin{figure}
    \centering
    \includegraphics[width=1.0\linewidth]{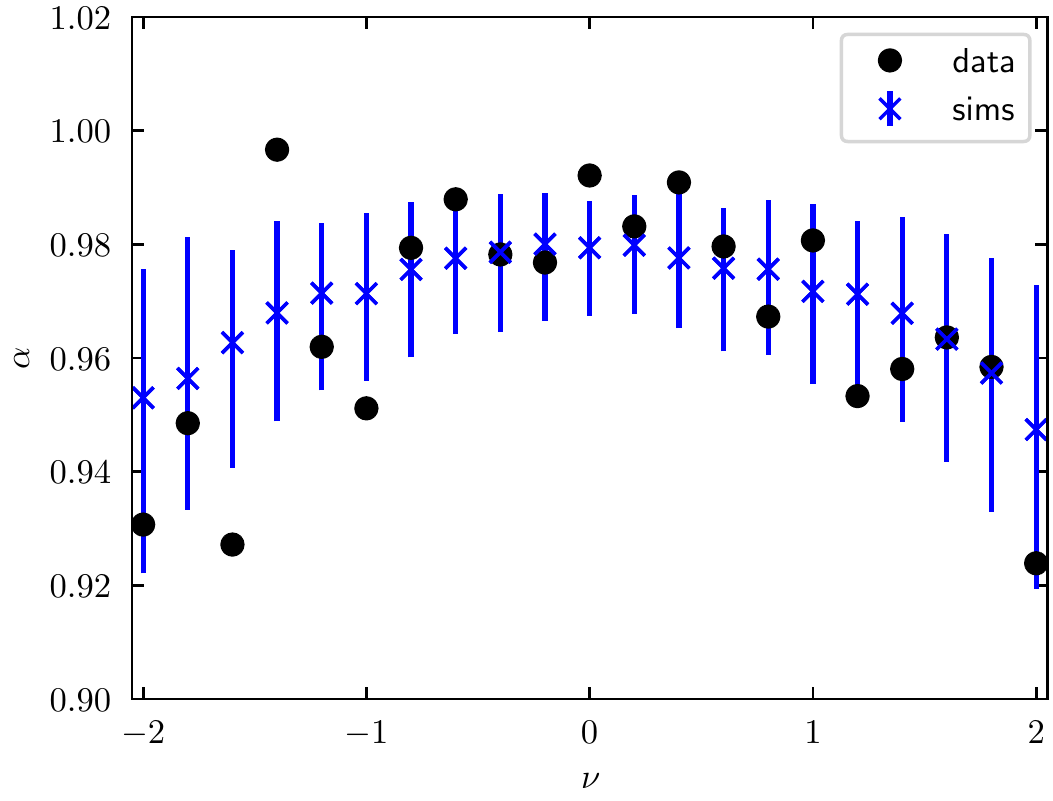} 
    \caption{$\alpha$ values from the data $E$ map (black points) and median of $\alpha$ from 300 FFP10 simulations (blue crosses).}
    \label{fig:alpha_iso}
\end{figure}

First, we check the consistency of $\alpha$ values computed from the data $E$ map with those obtained from the 300 SMICA simulations, which also include instrumental systematics effects already added to the noise component, as mentioned in section \ref{sec:data_sims}.
We compute $\alpha$ from the smoothed maps at $N_{\rm side}=128$ after applying P78 mask. The results of our analysis of the SI of the $E$-mode maps are presented in figure ~\ref{fig:alpha_iso}. The $\alpha$ values computed from the data $E$ map are represented by black points. The blue crosses with error bars represent the median $\alpha$ and 68\% limits from 300 SMICA simulations. We find that the combined $p$-value for the data $E$ map based on the 300 SMICA simulations is $p=0.54$, indicating that the data $E$ map is statistically consistent with the SMICA simulations.

\begin{figure}
\centering
    \includegraphics[width=1.0\linewidth]{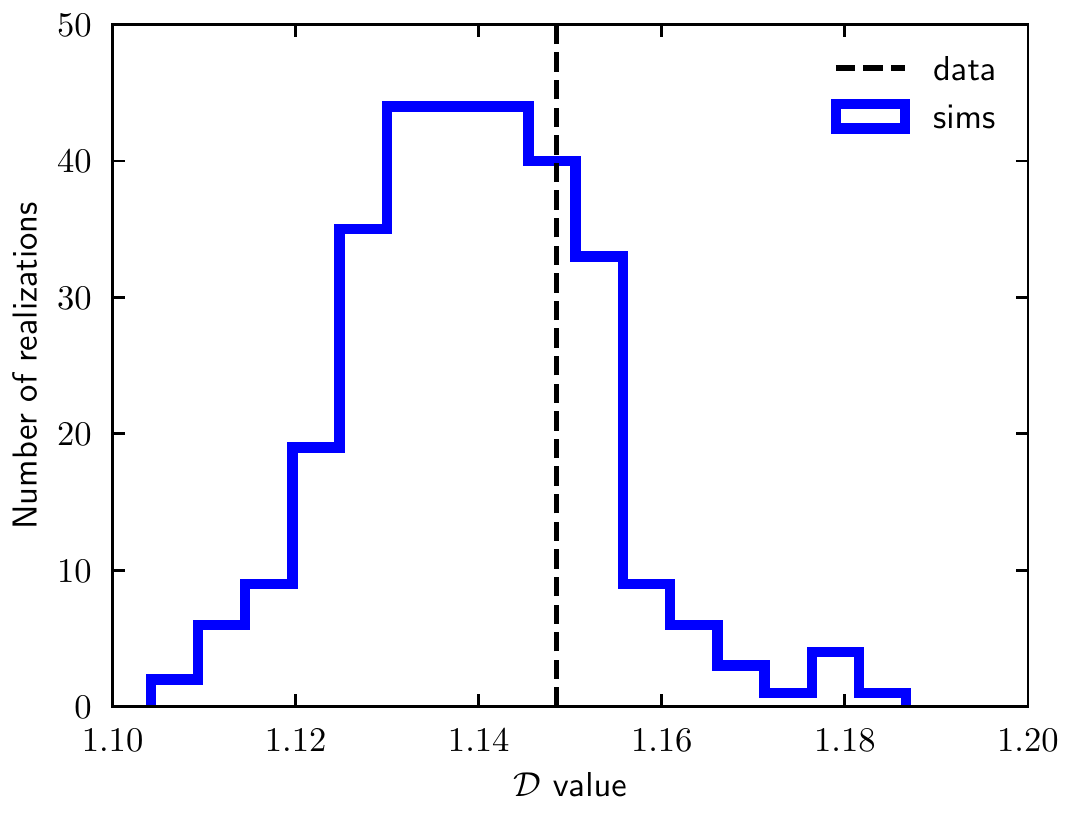}    
    \caption{The $\mathcal{D}$ value obtained from data $E$ map (dashed vertical line) and the histogram of 300 SMICA simulations (blue colour) computed over the masked sky.}
    \label{fig:D_comp}
\end{figure}

\vspace{0.1in}
Next, we present the SI results obtained using the $\mathcal{D}$ statistic. In order to check whether the CMB $E$-mode is SI or not, we have compared the $\mathcal{D}$ statistic results from the data $E$ map and SMICA simulations. The specifications of the data $E$ map and SMICA simulations are described in section ~\ref{sec:data}. We apply P78 mask to the Planck data and the simulations and the results of the $\mathcal{D}$ statistic over the masked sky are presented in figure ~\ref{fig:D_comp}. The $p$-value measured from the data $E$ map based on the 300 SMICA simulations is $p=0.23$. Our $\mathcal{D}$ statistic results do not detect any SI violation in the data $E$ map. This result concurs with earlier work done in \cite{ghrear2019testing}, where the authors have implemented $\mathcal{D}$ statistic on the CMB polarization angle map.

Our analysis using the $\alpha$ estimator and $\mathcal{D}$ statistic show that the data $E$ map is statistically consistent with the SMICA simulations.
As the input CMB $E$-mode signal in the SMICA simulations are SI, we  conclude that the CMB $E$-mode is statistically consistent with the SI approximation. This conclusion is supported by our demonstration in section \ref{sec:noisy} that the $\alpha$ estimator and $\mathcal{D}$ statistic are sensitive enough to pick up nSI signal in the presence of SI noise for $\gamma$ up to 1. For the data $E$ map, the effective $\gamma$ is slightly greater than 1. Hence, if there is a violation of SI in the CMB $E$-mode, similar to the one present in the filtered dust $E$-mode map used in section \ref{sec:noisy}, we expect to detect it using the $\alpha$ estimator and $\mathcal{D}$ statistic. It is important to note that the SMICA simulations used in this work to test the SI of the CMB $E$-mode contain only the CMB signal and the noise contributions propagated through the SMICA pipeline, with no residual foreground contamination. Hence, the non-detection of SI violation in the data $E$ map may be interpreted as evidence against significant levels of residual foreground contamination.
\section{Discussion}
We have applied the $\alpha$ estimator and the $\mathcal{D}$ statistic methods to the low resolution component separated SMICA $E$-mode map from the Planck 2018 data release, to test for any deviations from SI. We find that the data is consistent with the assumption of SI. Previously, the authors of \cite{ganesan2017tensor}, found a 4-$\sigma$ deviation from SI in the Planck 2015 $E$-mode data. They compared the Planck 2015 component separated $E$-mode maps with the 44GHz and 70GHz simulations. While it is not ideal to compare the component separated maps with individual frequency simulations, the authors at the time were limited by the availability of FFP simulations processed through the same component separation pipeline. The stereographic projection can introduce errors in the computation of $\alpha$ and this error was estimated and was shown to go as high as 24\% in Table 2 of \cite{ganesan2017tensor}. Our analysis is an improvement over the results in \cite{ganesan2017tensor}, in two ways. Firstly, we use the method of computing $\alpha$ directly on the sphere, avoiding the errors arising due to the stereographic projection. Secondly, we use the SMICA simulations, which are better suited for comparison with the component separated data maps, than the individual frequency simulations. The SMICA simulations which we use in our analysis, also have reduced noise levels and better modelling of systematics as compared to the corresponding simulations in the 2015 release.

\section{Conclusions}
\label{sec:conc}

The $\alpha$ estimator and $\mathcal{D}$ statistic provide independent tests of the SI of random fields in real space. For a given random field, $\alpha$ quantifies the level of alignment of structures in the level sets of the field. How far $\alpha$ lies away from unity towards zero, quantifies the level of alignment of the structures in the field, and hence is a measure of the deviation from SI in the field. Similarly, for $\mathcal{D}$ statistic the preferred directionality in the field is measured in terms of the alignment of the gradient vectors defined for the field. For an SI map the $\mathcal{D}$ value should be close to unity. Deviation from unity signifies the presence of statistical anisotropy. For a pixelated map, $\alpha$ is never equal to one even for an SI map. For this reason, we compare the $\alpha$ values computed numerically from observed maps with those computed from SMICA simulations which are SI. In this work, we have applied the CMT and $\mathcal{D}$ statistic techniques to the data $E$ map. We compare the $\alpha$ and $\mathcal{D}$ statistic values computed from the data $E$ map with those computed from SMICA simulations. The main results of the data analysis are as follows:

\begin{itemize}
    \item We find that the data $E$ map is statistically consistent with SMICA simulations based on the results obtained using the two estimators - $\alpha$ and $\mathcal{D}$ statistic. Since the input CMB $E$-mode signal in SMICA simulations are SI, we can conclude that the CMB $E$-mode polarization is statistically consistent with SI approximation.
    
    \item We test the sensitivity of the $\alpha$ estimator and $\mathcal{D}$ statistic for low signal-to-noise case. We add different levels of SI white noise to a nSI dust signal map and check the level of noise at which our two estimators are sensitive enough to pick up the input nSI dust signal. We find that with the addition of SI white noise to the original nSI signal, the resultant map becomes SI for very low noise levels ($\gamma<$ 1). For data $E$ map, the signal-to-noise ratio in the map space is marginally greater than 1. Our SI estimators are sensitive enough for the level of noise present in the data $E$ map.
    
\end{itemize}

\acknowledgments
Joby would like to thank NISER Bhubaneswar for the institute postdoctoral fellowship. All the Minkowski Tensor computations in this paper were run on the Aquila cluster at NISER supported by Department of Atomic Energy of the Govt. of India. TG acknowledges support from the Science and Engineering Research Board of the Department of Science and Technology, Govt. of India, grant number \texttt{SERB/ECR/2018/000826}. P. C. acknowledges support from the Science and Engineering Research Board of the Department of Science and Technology, Govt. of India, grant number \texttt{MTR/2018/000896}. AS would like to thank Majd Ghrear for some helpful discussions on $\mathcal{D}$ statistic. AS also acknowledges the use of Padmanabha cluster at IISER-TVM. Some of the results in this paper have been derived using the HEALPix ~\cite{gorski2005healpix,hpx} package. We acknowledge the use of the CMB polarization data provided by the Planck mission which is funded by the ESA member states, NASA, and Canada.

\appendix
\section{\texorpdfstring{$\mathcal{D}$}{D} statistic analysis of SMICA noise simulations}
Here we present the $\mathcal{D}$ statistic analysis of the SMICA noise-only simulations. Since the SMICA noise-only simulations include the residual systematics which may have a preferred direction, we expect the $\mathcal{D}$ statistic to pick this up. We apply the same post-processing to the noise-only simulations as we applied to the data in section \ref{sec:data_mask}, to produce noise-only simulations at $N_{\rm side}$=128 with Gaussian beam smoothing of $1^{\circ}$ FWHM. Next, we apply the binary P78 mask to these simulations and then run the $\mathcal{D}$ statistic analysis on them.

The median of $\mathcal{D}$ values from the 300 noise-only simulations is $1.281\pm 0.014$. We compare this value with those obtained from the SI simulations shown in figure \ref{fig:353fg+wnoise}, and find that there is no overlap between the two. We conclude that the SMICA noise-only simulations have directionality, based on the $\mathcal{D}$ statistic.

\bibliographystyle{unsrt}


\end{document}